# EXPECTED CASH FLOW:
# A NOVEL MODEL OF EVALUATING FINANCIAL ASSETS

Magomet Yandiyev[1]
Moscow State University, Economics Faculty
mag2097@mail.ru

*Abstract*: The present paper provides the basis for a novel financial asset pricing model that could avoid the shortcomings of, or even completely replace the traditional DCF model. The model is based on Brownian motion logic and expected future cash flow values.

*Keywords*: Cash Flow, Discounted, Expected, Present Value, Future Value, Islamic Finance

*JEL Classification*: G24

### 1. Introduction

The present paper aims to lay the foundation of a novel asset pricing model that would avoid the shortcomings of the universally recognised Discounted Cash Flow model (J. Williams, 1938), superseding the latter.

The Discounted Cash Flow model (DCF) works good for explaining price formation for bank credits and bonds, but proves substandard when used for stock and derivatives valuation and, in case of Islamic finance, totally unusable, credit and bond valuation included. One of the model's most commonly criticised aspects is that not all companies are planning on paying dividends within the planning horizon (D. Gode and J. Ohlson, 2006).

Arguably the most essential criticism of the model, nevertheless, is that DCF is based on the premise that there is an alternative of investing in a bank deposit available, and that, accordingly, these investments are equivalent. However, this premise is only valid for a very narrow set of circumstances. Consider a fairly typical example: there are a set number of investors and banks on the market. Each investor has an alternative way of investing his savings (into an investment project, stocks, bonds, or a bank deposit after all) and therefore thinks that DCF is applicable to evaluating the cost of this project. Suppose something makes these investors to simultaneously choose to exercise their right to an alternative investment and put all their money into banks. Responding to the surge in offer, the latter would lower the rate of return on deposits. Consequently, all prior DCF-based calculations involved in the investment projects (which assumed investing into a bank deposit would have an equal value) would prove incorrect.

Another criticism of DCF is that its premise of immediate reinvestment of dividends at the same rate is unrealistic and, in fact, impossible to achieve.

---

[1] Personal page on SSRN: www.ssrn.com/author=1278584

Moreover, it must be said that it is difficult to guess the future return on an investment, something that would define future cash flow discounts, because year on year, period on period it could fluctuate. *Guess*, because no model can predict the value.

That being said, the difficulty in predicting future cash flow should not be taken for a shortcoming exclusive to DCF. Any model based on future cash flow would be equally exposed to it.

Apart from these drawbacks, when it comes to Islamic finance, DCF has one more, fundamental flaw: it is based on loan interest, thus reducing its applicability to zero. Consequently, it is impossible to use DCF for evaluating assets within the framework of Islamic finance. It is DCF itself that needs an alternative.

## 2. Arguments for the ECF model

Suppose there is a financial asset that is promised to return a certain sum of money – sometime in the future, so there is always a degree of uncertainty about it. This uncertainty debases the future value of the promised sum by the degree of that uncertainty.

Given all that, the starting point of drawing up a model unaffected by all these shortcomings would be to premise that the present value of any future return is determined by the uncertainty of this return.

Then, for any financial instrument with the two known values – the sum expected to be paid in the future and the present value, discovered through market or off-exchange trading – the correlation between the values would be determined by the probability of that instrument's issuer discharging his financial commitments.

E.g., for an asset currently worth $100 and set to yield $110 in a year's time, its future value would be the correlation between the two prices adjusted for the probability of not receiving the sum. Thus, $100 is the expected value of $110.

To reiterate, the present value of an asset is a function of its future value:

$$PV = f[FV] \qquad (1)$$

where

- **PV** is the present value of a financial asset;
- **FV** is the future value of a financial asset.

The logic behind Equation (1) is nothing new. There are many asset pricing models based on expected value (Cochrane, 2000). In general, they are all addressing the same question: how to pinpoint the value of an asset. Depending of what factor is used for discounting, these models fall into two camps: consumption-based models and alternative models.

Consumption-based model's underlying formula is as follows:

$$PV_t = f_t\left[\beta \frac{u'(c_{t+1})}{u'(c_t)} FV_{t+1}\right] \qquad (1.1.)$$

with the discounting factor as $m_{t+1} = \beta \frac{u'(c_{t+1})}{u'(c_t)}$

where

- **t** is the moment in time;
- **β** is the subjective factor for adjusting for investor risk;
- **u'(ct)**, **u'(ct+1)** are derivatives of investor consumption functions;
- **ct**, **ct+1** are investor consumption at **t**, **t+1**.

Equation (1.1) is based on the idea that the investor discounts returns according to his preferences (marginal utility). The value is calculated to be the expected value of the product of marginal utilities (present to future) ratio and future returns on the asset. The model assumes the discounting factor to be variable, changing depending on the economic period. In a recession, present consumption is low, so marginal utility is high, which makes the ratio small, so asset value is likewise low.

Alternative models expand the discounting factor $m_{t+1}$ as a linear combination of various variables, with the most widely known example of a factor model being the CAPM.

The next (second) step in formulating the proposed new model would be asserting that the probability of reaching or not reaching a certain value is best described by the classical Brownian equation (for a 2D space):

$$S = \sqrt{n} * \lambda \qquad (2)$$

where

- **S** is displacement, i.e. the linear distance from the starting point of a particle's travel path to its end point;
- **n** is the number of times it moved or changed direction;
- **λ** is the average distance of one movement.

In the world of multiple economic, social, political, and other factors influencing the issuer of a financial asset much like the molecules bombarding the Brownian particle, applying the logic of Equation (2) seems only reasonable. All the more so given the fact that the idea of applying Brownian logic to asset pricing is not an altogether new one, and is already successfully used in the option pricing model (Black and Scholes, 1973).

The third step in formulating the proposed model would be replacing the original parameters of Equation (2) with their financial counterparts.

Equation (2) tells how many times (**n**) the Brownian particle changed its direction before transforming *average distance of one movement* (**λ**) into *displacement* (**S**). Consequently,

- *number of times the particle changed direction* (**n**)'s counterpart is the time period between **PV** and **FV**;
- *average distance of one movement* (**λ**) is the initial price of the asset, or **PV**;

- *displacement* (**S**) is the future value of the asset, or **FV**.

Therefore, Equation (2) can be rewritten as

$$FV = \sqrt{n} * PV \qquad (3)$$

The fourth step. Given that the Brownian equation measures **n** in the number of movements, while finance calculations usually deal with yearly periods, the contents of **n** need further clarification. Wuth the minimum period of time in a financial deal equalling one day, **n** is proposed to be measured in years, but accurate to a day:

$$n = \frac{X}{365} \qquad (4)$$

where

- **X** is the number of days between **PV** and **FV**;
- 365 is the number of days in a year.

However, in Equation (2) **n** is always greater than 1, hence the following correction:

$$n = \frac{X}{365} + 1 \qquad (5)$$

This way, Equation (3) becomes

$$PV = \frac{FV}{\sqrt{\frac{X}{365} + 1}} \qquad (6)$$

The resulting Equation (6) represents the natural probability similar for all issuers. However, different companies have different factors of safety against random events, and can enhance or, on the contrary, deteriorate their ability to discharge its obligations with, respectively, more efficient business organisation or otherwise. This calls for appending Equation (6) with a factor representing individual issuer characteristics:

$$PV = \frac{FV}{\sqrt{\frac{X(1-k)}{365} + 1}} \qquad (7)$$

where

- k is asset issuer efficiency, in fractions; a factor defining the issuer's ability to reduce natural payment default risks.

A simple illustration of Equation (7): an asset set to return $110 in a year's time is presently traded at $100. Applying Equation (7) gives the asset issuer's **k** to be 79%. This is a reasonably high value, signifying a high degree of certainty that it will discharge the promised payment. In other words, for this **k** the expected value of $110 at the present is $100.

Here **k** is given in per cents, even if in Equation 7 it is measured in fractions, ranging from 0 (satisfactory) to 100% (optimal). When **k** equals 0, it means that in securing the payment the issuer is utterly unable to reduce natural risks. When **k** equals 100%, it means the issuer's business is organised so efficiently that there is zero default risk (something practically impossible). That is, ranging from 0 to 1 **k** represents for the increase in probability of return. When **k** is negative, the issuer's ability to secure a return on the asset is below natural, though it does not imply complete insolvency.

By the logic of Equation (7), **k** can also be interpreted as the number of days the fund receiver (asset holder), in a way, saves though the efficiency of the issuer's business. In other words, a value of **k**<1 for a company means that, granting a loan for a duration of 365 days, a bank effectively assumes a risk equivalent to a shorter duration, namely **k***365 days.

Since it is based on the technique of expected values, it has been named the Expected Cash Flows (ECF) model. The model is applicable to credit, stocks, bonds, and annuity valuation.

### 3. Some characteristics of the ECF model

The model's discount factor behaves fairly logically. It shows a sharp decline of discounting factor for the short term and a more gradual lowering for the mid- and the long term. This agrees well with the nature of default risk, changing substantially if loan duration changes from, for example, 1 day to 1 month, but staying essentially the say if the duration changes from, for example, 3 years to 5 years.

The below Figure shows four lines, one plotting the discount factor according to ECF (discarding individual issuer characteristics, i.e., according to Equation (6)), and the other three plotting DCF discount factors at discounting rates of 5%, 10%, and 15% per annum:

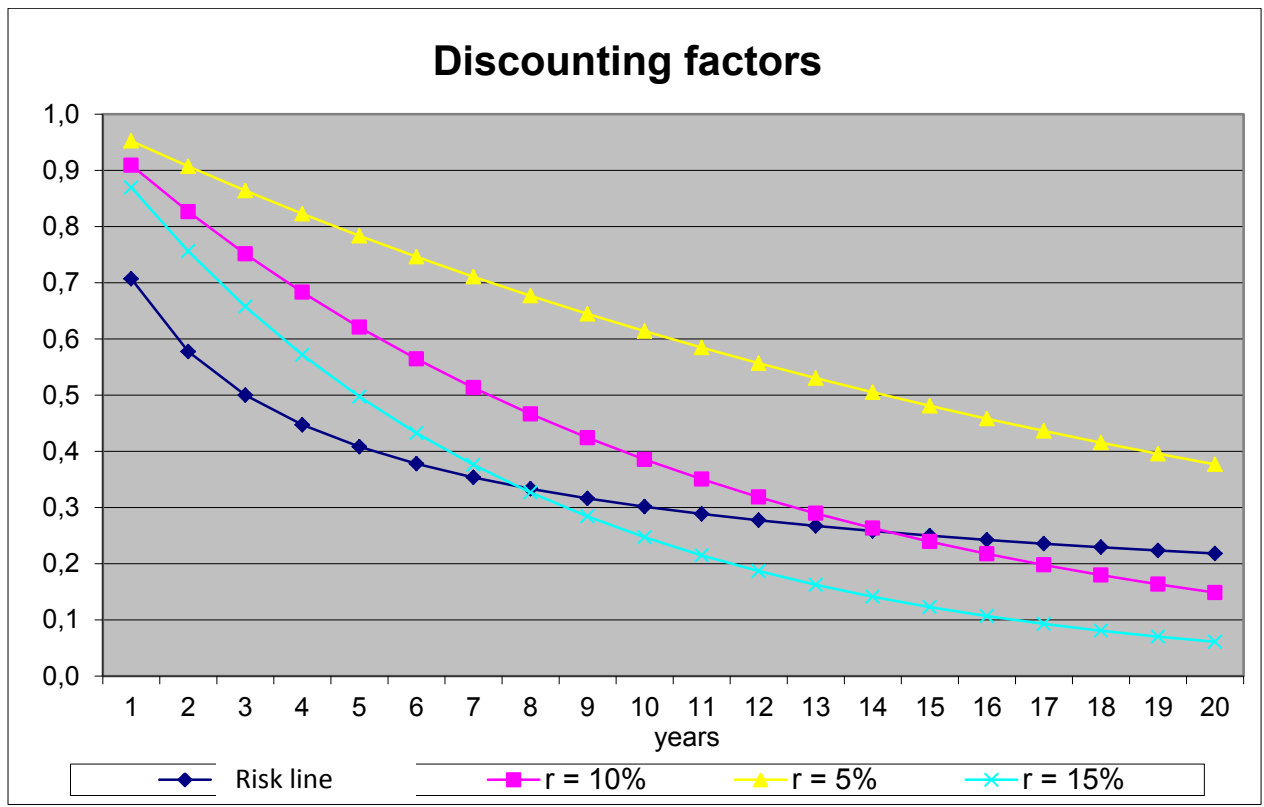

There are two further important aspects to ECF.

Firstly, for securities issued by the same company but with different maturity dates, **k** would not be the same. The longer the period, the lower **k** would be. In the following example, for a one-year asset **k** equals 0.79, for a two-year asset it falls to 0.77, and for a 15-year asset to 0.39 (see the following schedule; vertical: **k**, horizontal: years to maturity date),

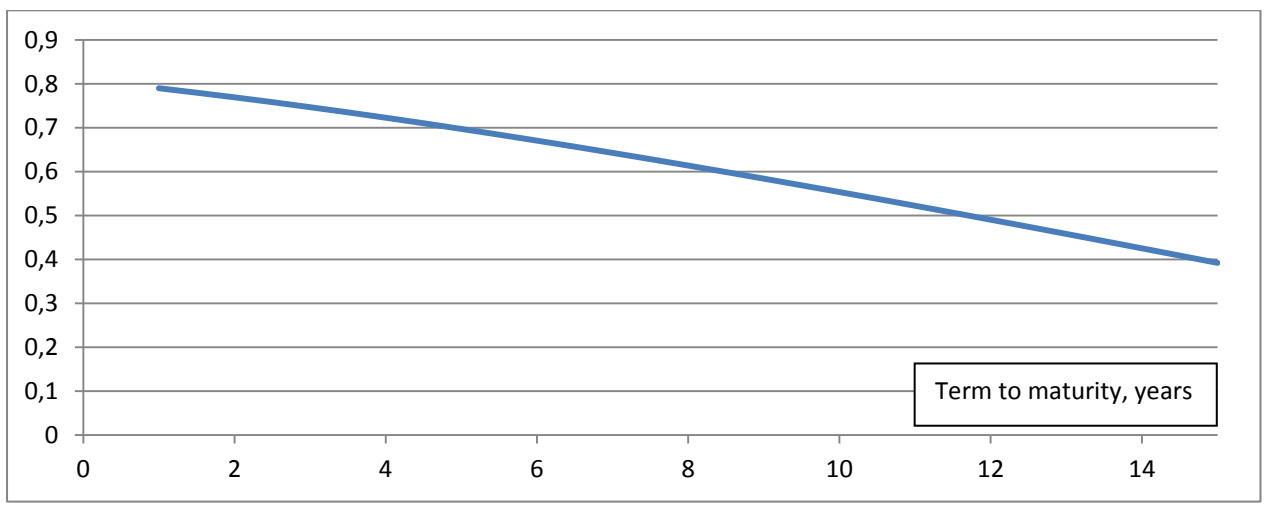

Secondly, however the market rate might change, an asset's present value would depend only on the market's perception of its issuer's individual characteristics. This is an entirely different model of reacting to events.

This calls for pointing out the degree of susceptibility to asset speculation as a shortcoming of the ECF model: faced with a speculative activity, its results, including the **k** factor, can become distorted. Consequently, for efficient application of ECF opportunities for speculation must be kept to a strict minimum.

4. **Conclusions**
   1) The ECF model helps avoid the main shortcomings of DCF: the erroneous premise of an alternative investment, and the necessity of guessing the future discounting rate. Moreover, ECF brings the number of basic defining parameters of cash flow down to just one (the **k** factor), against the whole set employed in DCF (WACC, CAPM, ATR, market index, bank deposit, etc.)
   2) ECF allows evaluating issuing company's business efficiency, defining its ability to discharge its obligations. At **k**=0, a company is totally incapable of reducing the risks inherent to the natural environment. At **k**=1, a company's business is said to be organised so efficiently that there is zero default risk, meaning **PV=FV.** This is not practically impossible, so for the best companies **k** simply tends to 1. The situation where a certain **k** value signifies a company being totally insolvent is not explicitly set, with the factor able to assume indefinite negative values.
   3) ECF as a model is applicable to Islamic finance. Moreover, the proposed **k** factor lays down the vision of Islamic financial scholarship's future development: this is a basis for organising the currency circulation and the interbank credit market in accordance with the principles of Sharia law. It should be noted that, when it comes to bank-client relationships, the techniques of Islamic finance are developed well enough. However the principles of interbank and central bank-commercial banks relationships are not yet thoroughly tuned. The consequence of which is that Islamic finance exists more as an adjunct of traditional finance rather than an independent self-reliant industry. This is set to go on unless the basic criteria for evaluating risk and return in Islamic finance are formalised; figuratively speaking, unless the Islamic world gets its own rate, as authoritative and recognisable as Libor.

In general, ECF proposes a novel model of thinking where asset value is independent of market price factors, inflation, or the number of interest calculations per year. Industry players would have to abandon the concept of interest rate and forget the DCF scenarios of asset behaviour.